\documentclass[aps,prx,superscriptaddress,twocolumn]{revtex4-1}
\usepackage{graphicx}
\usepackage{amssymb}
\usepackage{amsmath}
\usepackage{multirow}
\usepackage{float}
\graphicspath{{images/}}
\usepackage{dcolumn}  
\usepackage{bm}        
\usepackage{xcolor}
\begin{document}

\title{Observation of superconducting gap spectra of long-range proximity effect in Au/SrTiO$_3$/SrRuO$_3$/Sr$_2$RuO$_4$ tunnel junctions}

\author{M.~S.~Anwar}\email[E-mail:]{msa60@cam.ac.uk}\affiliation{Department of Physics, Kyoto University, Kyoto 606-8502, Japan}\affiliation{Department of Materials Science and Metallurgy, University of Cambridge, Cb40FS, Cambridge, United Kingdom}

\author{M.~Kunieda}\affiliation{Department of Physics, Kyoto University, Kyoto 606-8502, Japan}

\author{R.~Ishiguro} \affiliation{Department of Mathematical and Physical Sciences, Faculty of Science, Japan Women's University, Tokyo 112-8681, Japan} \affiliation{Department of Applied Physics, Faculty of Science, Tokyo University of Science, Tokyo 162-8601, Japan}

\author{S.~R.~Lee}  \affiliation{Center for Correlated Electron Systems, Institute for Basic Science (IBS), Seoul 151-747, Republic of Korea}\affiliation{Department of Physics and Astronomy, Seoul National University, Seoul 151-747, Republic of Korea}

\author{L.~A.~B.~Olde~Olthof}\affiliation{Department of Materials Science and Metallurgy, University of Cambridge, Cb40FS, Cambridge, United Kingdom}

\author{J.~W.~A.~Robinson}\affiliation{Department of Materials Science and Metallurgy, University of Cambridge, Cb40FS, Cambridge, United Kingdom}

\author{S.~Yonezawa} \affiliation{Department of Physics, Kyoto University, Kyoto 606-8502, Japan}

\author{T.~W.~Noh} \affiliation{Center for Correlated Electron Systems, Institute for Basic Science (IBS), Seoul 151-747, Republic of Korea}\affiliation{Department of Physics and Astronomy, Seoul National University, Seoul 151-747, Republic of Korea}

\author{Y.~Maeno} \affiliation{Department of Physics, Kyoto University, Kyoto 606-8502, Japan}

\date{\today}

\begin{abstract}

We observe an unconventional superconducting minigap induced into a ferromagnet SrRuO$_3$ from a spin-triplet superconductor Sr$_2$RuO$_4$ using a Au/SrTiO$_3$/SrRuO$_3$/Sr$_2$RuO$_4$ tunnel junction. Voltage bias differential conductance of the tunnel junctions exhibits V-shaped gap features around zero bias, corresponding to a decrease in the density-of-states with an opening of a superconducting minigap in SrRuO$_3$. Observation of a minigap at a surface of a 15~nm thick SrRuO$_3$ layers confirms the spin-triplet nature of induced superconductivity. The shape and temperature dependence of the gap features in the differential conductance indicate that the even-frequency $p$-wave correlations dominate, over odd-frequency $s$-wave correlations. Theoretical calculations support this $p$-wave scenario. Our work provides the density-of-states proof for $p$-wave Cooper pair penetration in a ferromagnet and significantly put forward our understanding of the $p$-wave spin-triplet proximity effect between spin-triplet superconductors and ferromagnets.
\end{abstract}

\maketitle

\section{Introduction}

Spin-triplet superconductivity is rich in physics due to its spin and orbital degrees of freedom compared to its counterpart spin-singlet superconductivity. It not only exists in bulk materials like Sr$_2$RuO$_4$ (SRO214), UPt$_3$, {\it etc.}~\cite{Maeno2012} but also emerges in superconductor-ferromagnet (SSC/F) heterostructures that exhibit particular broken symmetries~\cite{Bergeret2001,Keizer2006,Eschrig2008,Khaire2010,Robinson2010,Anwar2010,Anwar2012,Bernardo2015}. Devices based on spin-triplet superconductivity are the fundamental building blocks to generate dissipationless spin-polarized supercurrents required to established superconducting spintronics~\cite{Bergeret2005,Linder2015}. In the last two decades, extensive theoretical and experimental knowledge has been developed to understand the generation of spin-triplet correlations in SSC/F junctions. In such junctions, spin degree of freedom may not be fully preserved since SSC has zero spin polarization. This issue can be solved by replacing SSC with a spin-triplet superconductor (TSC). However, crucial concerns in using TSCs are, firstly, availability of handful amount of bulk TSCs, and secondly, their compatibility to form an electronically transparent interface with other materials. Recently, some of the present authors developed an epitaxial TSC/F heterostructure by growing ferromagnetic SrRuO$_3$ (SRO113) thin films on superconducting SRO214 substrates~\cite{Anwar2015}. Furthermore, long-range spin-triplet proximity effect induced into SRO113~\cite{Anwar2016} was observed. Now, it is highly required to study the symmetry of induced correlations.

Superconductivity occurs in SRO214 with superconducting critical temperature ($T_{\rm c}$) of 1.5~K. Extensive experimental and theoretical studies~\cite{Maeno2012} indicate that SRO214 exhibits a chiral $p$-wave spin-triplet state with spontaneous breaking of the time-reversal symmetry~\cite{Maeno1994,Luke1998,Ishida1998,Nelson2004,Xia2006,Xia2006,Kindwingira2006,Anwar2013,Anwar2017,Anwar2016}, although there are unresolved issues~\cite{Hicks2010,Yonezawa2013}. Furthermore, SRO214 has recently been attracted interest for exploring topological superconducting phenomena originating from its orbital phase winding~\cite{Maeno2012}.

Spin-triplet superconductivity at SSC/F interfaces exhibits various subgap features depending on the symmetry of the induced correlations. These subgap features can be observed in the electronic density of states (DoS)~\cite{Eschrig2003, Wang2013}. Recently, zero-bias conductance peaks (ZBCPs) corresponding to odd-frequency $s$-wave spin-triplet correlations were observed in various experiments using metallic~\cite{Bernardo2015,Pal2017} and oxide-based SSC/F systems~\cite{Kalcheim2012,Kalcheim2014,Kalcheim2015,Bernardo2017}. Bernardo {\it et al.}~\cite{Bernardo2017} reported the observation of $p$-wave correlations in graphene connected with a $d$-wave high temperature superconductor Pr$_{2-x}$Ce$_x$CuO$_4$ by scanning tunnelling microscopy (STM). They observed a variety of subgap structures such as V-shaped gaps, ZBCPs, and split ZBCPs, depending on the position of the STM tip. %Their observed spectra are consistent with theoretical calculations for induced $p$-wave superconductivity in graphene.

Recently, some of the present authors developed superconducting junctions based on SRO214 in combination with the itinerant ferromagnet SRO113, where direct penetration of superconducting correlations over a 15-nm-thick SRO113 layer was observed, through multiple Andreev reflection features in Au/SRO113/SRO214 proximity junctions~\cite{Anwar2016}. To confirm this long-range penetration and to investigate symmetry of the induced superconductivity, DoS measurements are required. Moreover, in that system, it was argued that $p$-wave correlations may dominate since the superconducting source was of $p$-wave and the SRO113 layer was thinner than electron mean free path ($l_e$).  This is a unique and interesting possibility, but experimental verification is still absent.

Here, to address these questions, we developed tunnel junctions by depositing a 2-nm-thick insulating SrTiO$_3$ (STO) layer between the F-layer SRO113 and Au electrode. We performed differential conductance measurements of the Au/STO/SRO113/SRO214 heterostructures. A V-shaped gap feature in the conductance spectra, corresponding to the superconducting minigap induced in a 15-nm-thick SRO113 layer, is observed. Previous studies suggest that $p$-wave spin-triplet proximity effect leads to such a V-shaped gap~\cite{Wang2013, Bernardo2017}. Furthermore, our theoretical calculations confirm the $p$-wave symmetry of the induced superconductivity in the SRO113 F-layer. 

\section{Experimentation}

Single crystals of SRO214 with minimal eutectic segregation of Sr$_3$Ru$_2$O$_7$, SRO113, and Ru are carefully selected at cost of slightly reduced $T_{\rm c}$ and utilized to fabricate Au/STO/SRO113/SRO214 junctions. Ferromagnetic SRO113 thin films are grown epitaxially by pulsed laser deposition on cleaved $ab$-surfaces of SRO214 substrates with a thickness of 0.5~mm and a surface area of 3$\times$3~mm$^2$ (Further details in Ref.~\cite{Anwar2015}). Immediately after the deposition of SRO113, a 2-nm-thick insulating STO layer followed by a Au(20nm)/Ti(5nm) capping layer, are deposited {\it ex-situ} by DC sputtering. 

\begin{figure}
\begin{center}
	\includegraphics[width=8cm]{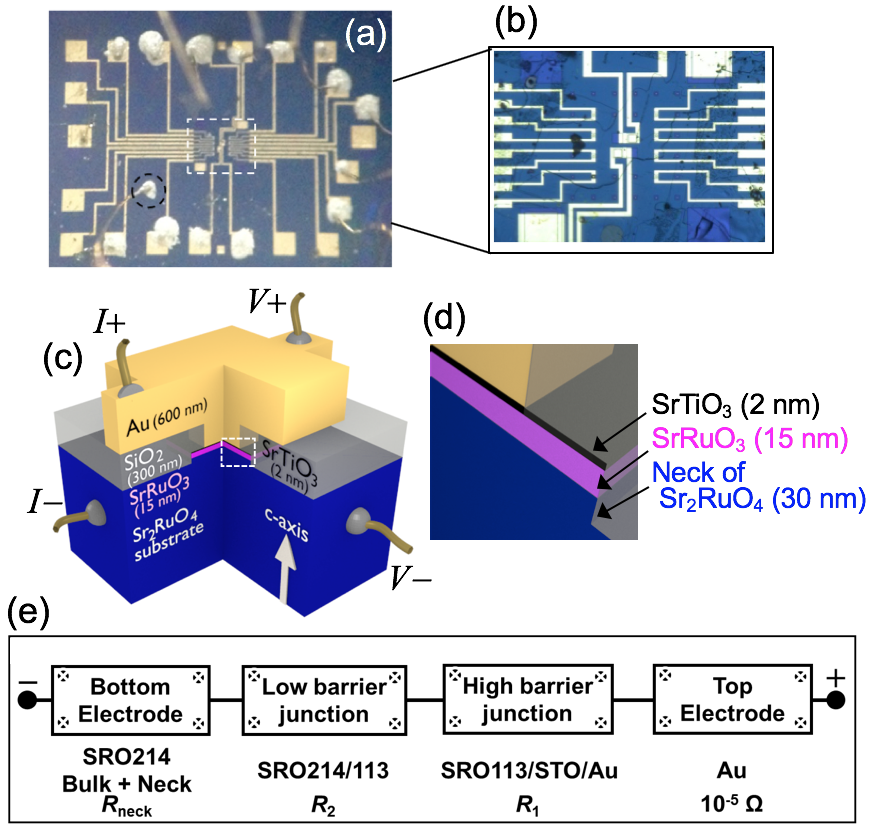}
	\caption{Au/STO/SRO113/SRO214 tunnel junctions. (a) Optical micrograph of junctions fabricated on a SRO214 substrate. (b) Magnified micrograph of device area indicated with a white dashed rectangle in (a). (c) Schematic three-dimenssional view of the junction. (d) Neck strucutre of SRO214 below SRO113. (e) Series resistor circuit model of the junction.}
	\label{device}
\end{center}
\end{figure}

Au/STO/SRO113/SRO214 tunnel junctions with areas 20$\times$20~$\mu$m$^2$ and 5$\times$5~$\mu$m$^2$ are fabricated on 25~$\times$~25 $\mu$m$^2$ and 10~$\times$~10~$\mu$m$^2$ SRO113 pads by laser UV maskless photolithography. The junction area i.e. the size of the top Au electrode, is smaller than the SRO113 pads to avoid contact between the top Au electrode and bottom SRO214 substrates (Fig.~\ref{device}). 

Electrical transport measurements are performed using a four-point technique with two contacts on the top Au electrode and two directly on SRO214 (Fig.~\ref{device}(c)). Resistivity and differential conductance are measured down to 300~mK using a ${}^3$He cryostat with a superconducting magnet. The bulk critical temperature $T_{\rm c-bulk}$ of the SRO214 single crystal was found to be 1.25~K (inset of Fig.~\ref{RT}(a)).  

\section{Results and discussion}

\begin{figure}
	\begin{center}
		\includegraphics[width=8cm]{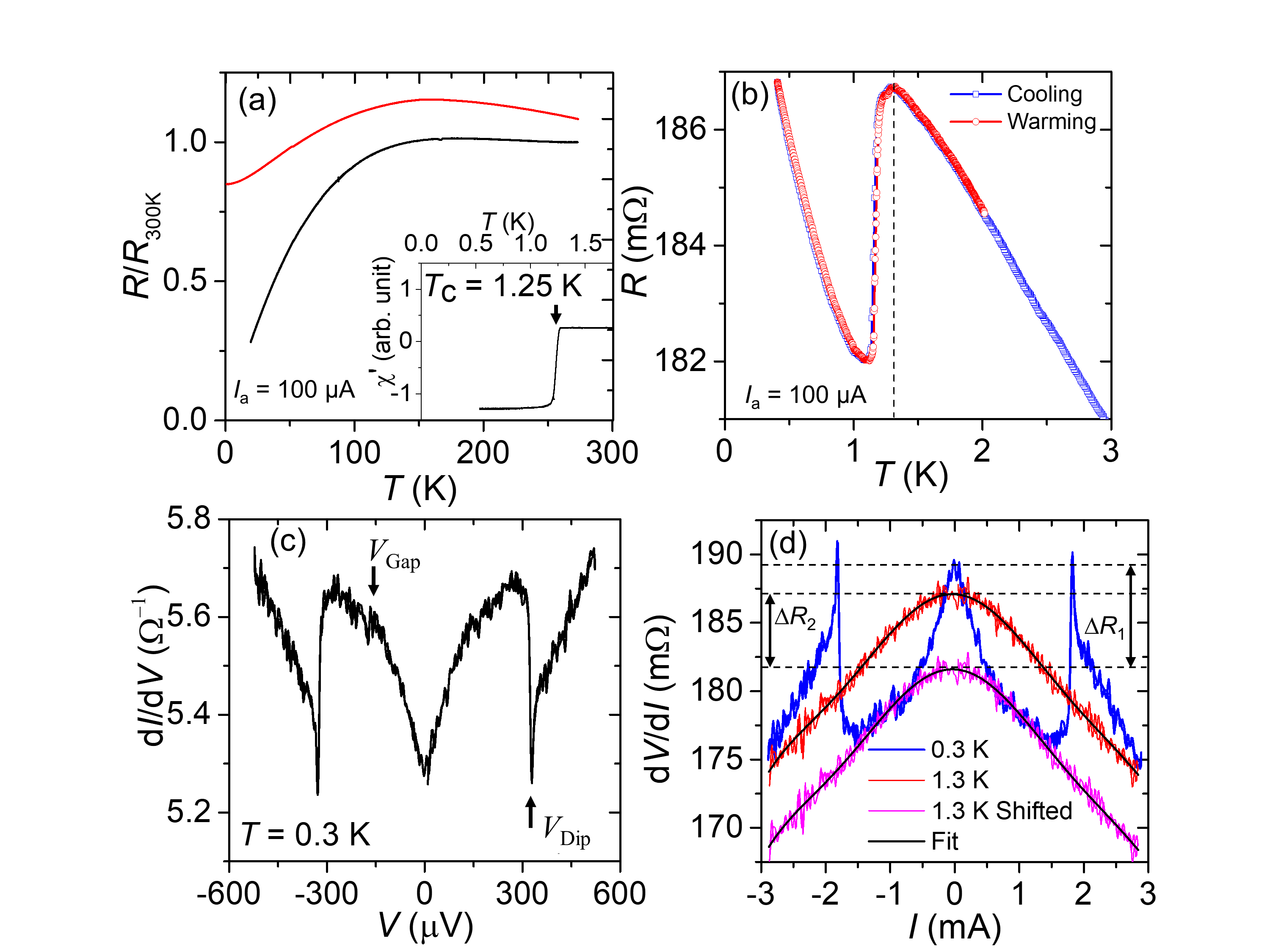}
\caption{(a) Temperature dependent resistance $R$($T$) measured at higher temperatures for junctions with (red) and without (black) the STO barrier. These sets of data show that, in this temperature range, $R_{c}$ of the SRO214 substrate dominates in both junctions. The inset shows that the AC magnetic susceptibility of the SRO214 substrate exhibits a sharp transition at $T_{\rm c-bulk}$ = 1.25~K. (b) $R$($T$) below 3~K measured with 100~$\mu$A current for the junction with an STO barrier. A sharp superconducting transition was observed below $T_{\rm c-bulk}$. Note that the resistance continuously increases with decreasing temperature $T_{\rm c-bulk}$ because of the STO tunnel barrier. (c) Differential conductance $dI$/$dV$ at 0.3~K, showing a V-shaped gap around the zero bias and two strong dips at higher voltages. (d) Differential resistance $dV$/$dI$ obtained at 0.3~K (blue curve) and 1.3~K (red curve) plotted as a function of applied current. A central gap opens up and two sharp peaks at 1.8~mA corresponding to critical current of the first part of the junction SRO113/SRO214. To analyse the effects of tunnel barrier only, $\Delta R_{2}$ $\approx$5.5~m$\Omega$ (resistance contributions of the neck and SRO113/SRO214 interface) is subtracted from the curve measured at 1.3~K (shifted magenta curve).}
		\label{RT}
	\end{center}
\end{figure}

Figure~\ref{RT}(a) shows the temperature-dependent resistance {$R$}($T$) in the normal state (300~K to 4~K) of a 5$\times$5~$\mu$m$^2$ junction (red curve). The resistance slowly increases  with decreasing temperature {\it T} down to 170~K, suggesting dominance of the $c$-axis bulk resistivity of SRO214 ($\rho_c = 1$~m$\Omega$cm at 4~K)~\cite{Hussey1998}. This behavior indicates that the current flows along the normal to the junction, and that direct electrical contact between Au and SRO214 is absent. With further decrease of  temperature below 100~K, $R$ does not decrease substantially, indicating dominance of the resistive contribution of the Au/STO/SRO113 tunneling junction. Consequently, the residual resistance ratio (RRR) is low (1.25). For comparison, $R$($T$) of a metallic junction (without STO layer) exhibits a RRR of 9, as shown with the black~curve. These observations demonstrate STO layer is working as a tunnel barrier.

At temperatures below 6~K, the $R$ increases with decreasing $T$ due to STO tunnel barrier. A sharp decrease of $R$ at 1.2~K is observed (Fig.~\ref{RT}(b)) corresponding to the superconducting transition of SRO214. With a further decrease in $T$, $R$ increases due to the superconducting gap opening in the superconducting state, since bias voltage $V$ is much smaller than the expected superconducting gap $\Delta$ of SRO214. Such $R$($T$) behavior was not observed in metallic junctions~\cite{Anwar2015, Anwar2016}, again indicating the tunnelling behavior of the present junction with the STO layer.

We now discuss the differential conductance  $dI/dV$ behavior of the tunnel junction. In Fig.~\ref{RT}(c), we have plotted $dI$/$dV$ at 0.3~K. Two main features are observed. The first sharp dips at $\pm$0.3~mV appears immediately below the $T_{\rm c}$ . Such dips mainly appears due to current-driven distraction of superconductivity in superconducting junctions~\cite{Sheet2004,Yang2012}. It most probably indicates that dips in $dI$/$dV$ data correspond to the critical current transition at SRO113/SRO214 interface, as we discuss later. The second but most important feature, is the suppression of conductance around the zero bias within $\pm$150~$\mu$V indicating a superconducting gap opening. This gap opening is due to the minigap of induced superconductivity in the SRO113 layer. We observed this behaviour in various junctions (see Supplemental Material~\cite{SM}). 

To extract the conductance spectra of the Au/STO/SRO113 junction more quantitatively, we estimated and subtracted the resistance contributions from the SRO214 neck ($R_{\rm neck}$) and SRO113/SRO214 interface ($R_2$) as follows. Our junction consists of series of various components, as depicted in the model shown in Fig.~\ref{device}(e). In such a series circuit, the differential resistance $dV/dI$ as a function of $I$ is appropriate for extraction of each contribution since $I$ is common to the all components and the resistance behaves additively. Thus, we plotted $dV/dI$ vs $I$ at 0.3 K and 1.3 K in Fig.~\ref{RT}(d). In the superconducting state, mainly the Au/STO/SRO113 tunnel junction contributes to the zero bias resistance. However, the resistance of the SRO113/SRO214 interface can have small but non-negligible contribution. In the normal state, the SRO214 neck under the SRO113 pad (see Fig.~\ref{device}(d)) adds an additional resistance due to higher resistivity of SRO214 along the $c$-axis~\cite{Anwar2015,Anwar2016}. We subtracted the contributions of the neck and SRO113/SRO214 by shifting d$V$/d$I$ curve at 1.3~K by $\Delta R_2 = R_2 + R_{\rm neck}=$5.5~m$\Omega$. The shifted curve is used to normalise the obtained conductance data at various temperatures and applied fields. More details are given in the Supplemental Material~\cite{SM}.

\begin{figure}
	\begin{center}
		\includegraphics[width=8cm]{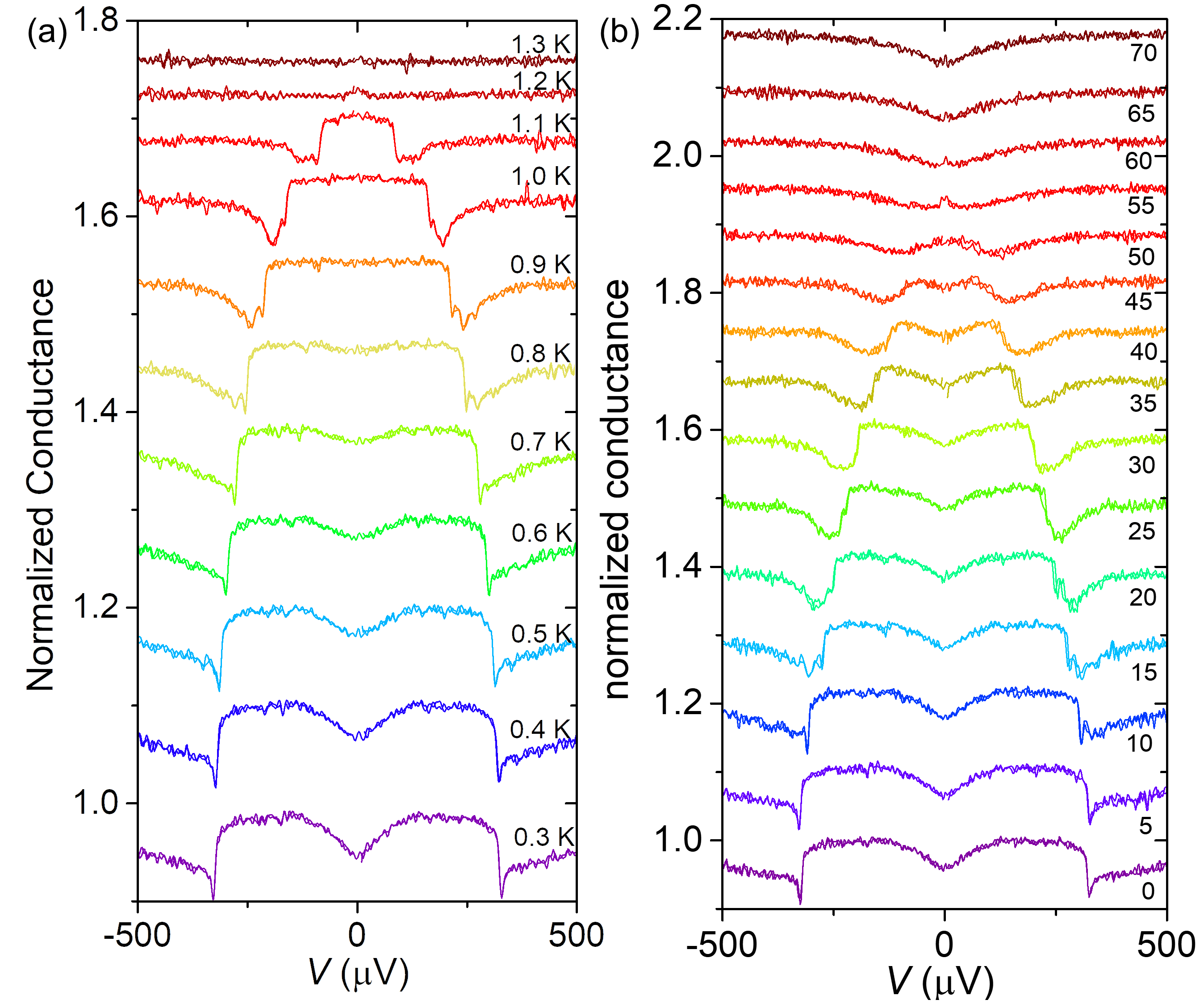}
\caption{(a) Normalized differential conductance as a function bias voltage obtained at different temperatures in zero field, and (b) at various applied out-of-plane magnetic fields in mT at 0.3~K.}
		\label{dIdV}
	\end{center}
\end{figure}

\begin{figure}
	\begin{center}
		\includegraphics[width=8cm]{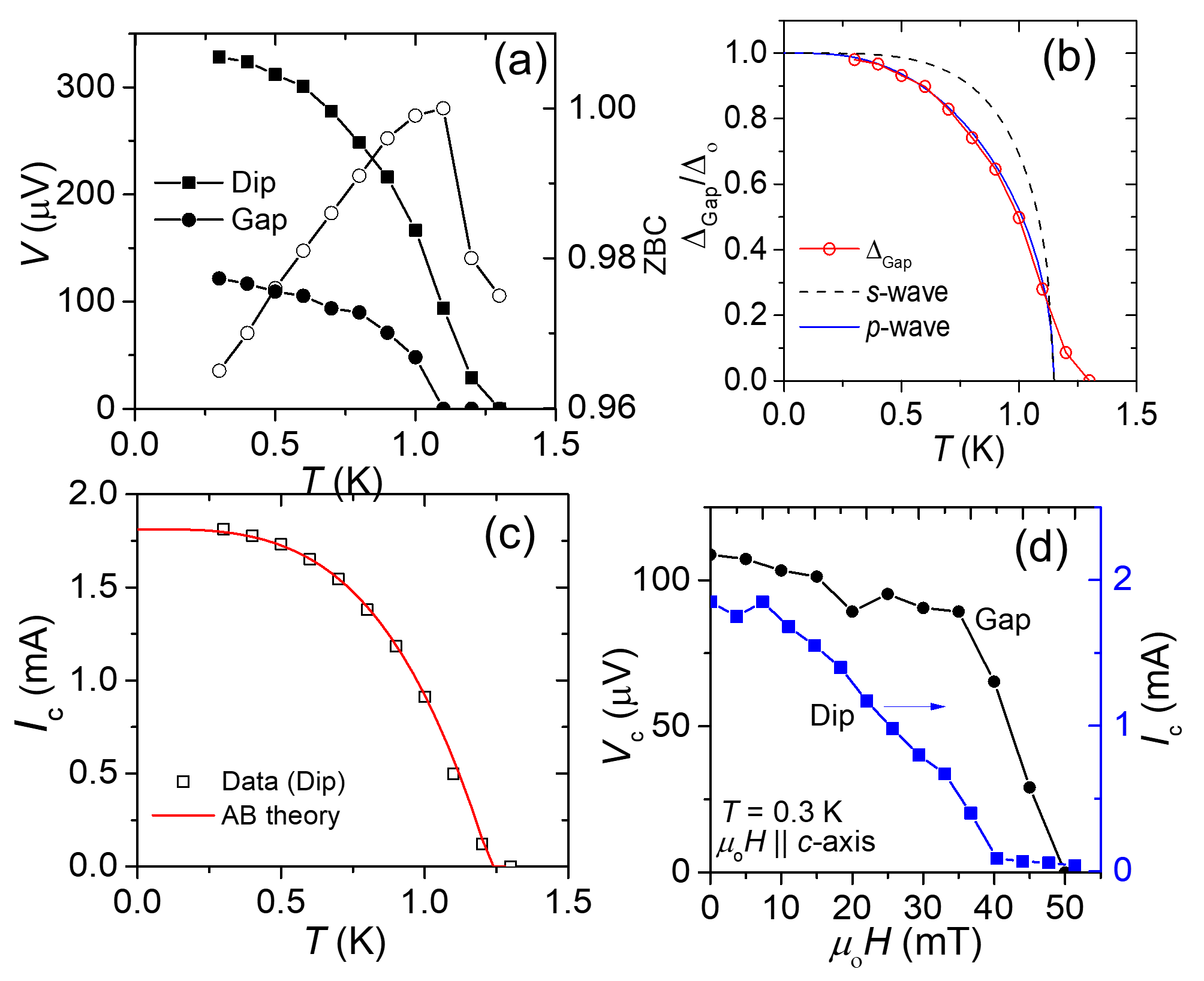}
\caption{(a) Minigap (filled circle) and dip (filled square) as well as normalized zero bias conductance (open circles) vs $T$. (b) Normalized value of minigap vs $T$ with different theoretical fits. (c) Critical current as a function of $T$ of SRO113/SRO214 junction. The AB theory shows a good fit (solid red line). (d) Field dependence of minigap and dip measured at 0.3~K.}
		\label{Analysis}
	\end{center}
\end{figure}

To understand the features in detail, we measured $dI$/$dV$ at various temperatures and magnetic fields applied along the $c$-axis (out-of-plane) (Fig.~\ref{dIdV}). Both $V_{\rm Gap}$ (characteristic voltage of central gap opening) and $V_{\rm Dip}$ (characteristic voltage corresponding to the sharp dips) become suppressed with increasing temperature or applied field. Furthermore, these features are observed only below $T_{\rm c}$, indicating that they originate from superconductivity in the junction. As shown in Fig.~\ref{Analysis}(a), $V_{\rm Gap}$ disappears above 1~K, well below $T_{\rm c}$. However, $V_{\rm Dip}$ survives to $T_c$ of SRO214. This different temperature dependence can be explained as follows: induced correlation is suppressed with the increase in the temperature and disappears first at the Au/STO/SRO113 interface and then at the SRO113/SRO214 interface. These observations suggest that $V_{\rm Gap}$ and $V_{\rm Dip}$ emerge from different interfaces of the junction; Au/STO/SRO113 and SRO113/SRO214, respectively. Figure~\ref{Analysis}(d) shows $V_{\rm Gap}$ and $V_{\rm Dip}$ as a function of applied field ($H$).

The suppression of the differential conductance around zero bias is due to the decrease in DoS because of superconducting minigap opening. This observation demonstrates that a proximity effect over 15~nm in to SRO113 and confirm the long-range spin-triplet correlations in SRO113, since the spin-singlet superconducting coherence length for SRO113 is only $\xi_{\rm F} = 1$~nm~\cite{Anwar2016}. 

The value of the minigap with $V_{\rm Gap}=150~\mu$V at 0.3~K decreases monotonically with increasing temperature. We here assume that the minigap is proportional to the bulk superconducting gap and consider the theoretical temperature dependence of the bulk gap, $\Delta(T)=\Delta_0{\rm tanh}(A\sqrt{\frac{T_{\rm c}}{T}-1})$, where $\Delta_0$ is the superconducting energy gap at $T=0$ and $A$ is a constant which is 1.74 for an $s$-wave BCS gap. Our data follow this relation with $A=1.56$ equivalently to the calculations of Nomura and Yamada~\cite{Nomura2002} for $p$-wave superconductivity (see Fig.~\ref{Analysis}(b)). Furthermore, $2\Delta_0/k_{\rm B}T_{\rm c}=3.2$ (Here we took $T_{\rm c}=1.1$~K since the gap feature disappears around 1.1~K), which is lower than the expected value of 4.3 for $s$-wave superconductivity. These parameter values also suggests the unconventionality of the induced minigap in the SRO113 layer. 

The dominating orbital symmetry of spin-triplet correlations at an SSC/F interface can be odd-frequency $s$-wave or even-frequency $p$-wave, depending on the thickness of the F-layer ($t_{\rm F}$) or length of the junction~\cite{Eschrig2008}. For a diffusive junction ($l_{\rm e}<t_{\rm F}$), odd-frequency isotropic $s$-wave spin-triplet correlations dominate since $p$-wave is sensitive to potential scatterings. In contrast, in a clean/ballistic junction ($l_{\rm e}>t_{\rm F}$), even-frequency anisotropic $p$-wave correlations may take over. In our junctions, the nature of the induced correlations can be $p$-wave spin-triplet because $t_{\rm F}=15$~nm is shorter than $l_e=20$~nm. A junction with a barrier between the top Au electrode and SRO113 layer probes the minigap of the induced superconductivity in SRO113. The shape of the observed minigap is V-shaped (see Fig.~\ref{RT}(c)), which supports the even-frequency anisotropic $p$-wave scenario. 

A SRO113 layer with residual conductivity $\rho_0 =10~{\rm \mu}\Omega$cm exhibits an electron mean free path of $l_{\rm e} \approx 20$~nm, which is larger than the thickness of SRO113 used in our junctions. We can therefore consider that our junctions are in the clean limit. The bias voltage corresponding to the minigap should be of the same order as the Thouless energy for a clean system (the escape energy corresponding to inverse of the escape time) $E_{\rm Th}=\hbar v_{\rm F}/t_{\rm F}$, where $v_{\rm F} \approx 1 \times 10^{6}$~cm/s is the Fermi velocity~\cite{Alexander2005} and $t_{\rm F}=15$~nm is the thickness of the SRO113 layer. It leads to $E_{\rm Th} \approx 440~{\rm \mu}$eV, which is of the same order as the measured minigap value $V_{\rm c} \approx 150~{\rm \mu}$V at 0.3~K. For comparison, we can estimate $E_{\rm Th}$  in diffusive limit by using shorter $l_{\rm e} \approx 10$-nm and using the formula $E_{\rm Th} = \hbar D/t_{\rm F}^2$, where $D$ is the diffusion coefficient. The diffusion coefficient can be calculated using free electron model $e^2\rho_0DN=1$, where $e$ is the electron charge and $N$ is the density of states of SRO113 at its Fermi level. Using $\rho_0\approx 30~\mu\Omega$cm and $N\approx 1 \times 10^{47}$~states/Jm$^3$, gives $D \approx 13$~cm$^2$/s and thus $E_{\rm Th} \approx $~4~meV, which is an order  of magnitude higher than that of the clean limit and the obtained  value for the minigap. These comparisons support that our junctions are in the clean limit. In such a system, an even-frequency $p$-wave superconducting order parameter may dominate over the odd-frequency $s$-wave spin triplet.

\begin{figure}
	\begin{center}
		\includegraphics[width=8cm]{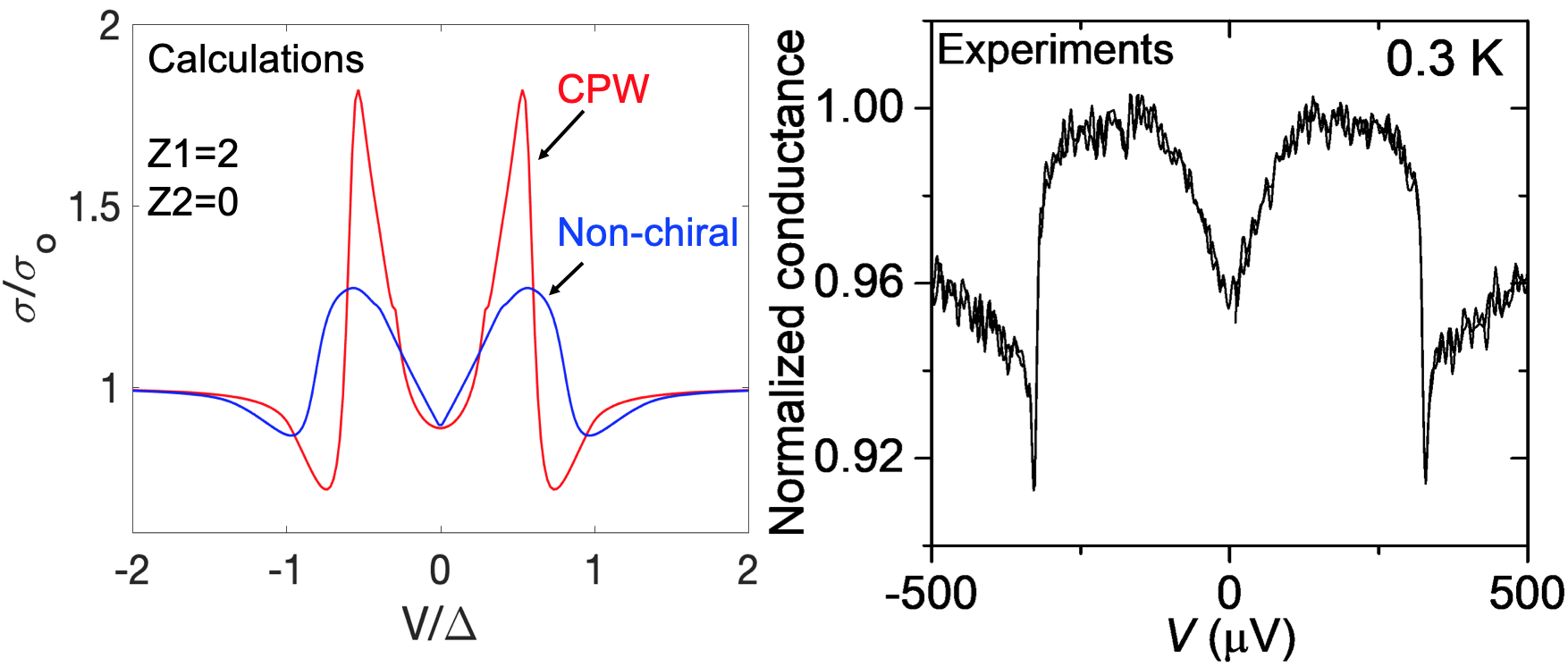}
\caption{(a) Calculated differential conductance vs biase voltage for chiral $p$-wave and non-chiral $p$-wave superconducting correlations with $Z_1=2$, and $Z_2=0$. (b) Normalized differential conductance measured at 0.3~K.}
		\label{Calculations}
	\end{center}
\end{figure}

To understand the nature of the induced minigap, we calculate the differential conductance of a Normal-metal(N)/Insulator(I)/F/TSC junction based on a recent theoretical model~\cite{Linde2018}. This model was developed particularly for Au/SRO113/SRO214 junctions, but can be applied to our system by considering that the barrier heights $Z_1$ and $Z_2$ correspond to the Au/STO/SRO113 and SRO113/SRO214 interfaces, respectively. We assume the $p$-wave order parameters for induced superconductivity in the SRO113 layer, $\Delta=\Delta_0 (k_x+i\chi k_y)\hat{\sigma}_x$, where $\Delta_0$ is the superconducting energy gap at $T=0$, $\sigma_x$, $\sigma_y$, and $\sigma_z$ are Pauli matrices and $\chi=\pm 1$ is the chirality. For normalization purposes, the superconducting gap is defined as $\Delta_0=1$. We assume the chemical potential is constant across the junction and equal to $\mu = 1000\Delta_{o}$.

The effective masses of the F and TSC layers are normalized with respect to the mass of the N layer as $m_{\rm N}=1$, $m_{\rm F} = 7m_{\rm N}$ (SRO113) and $m_{\rm S}^\parallel=1.3m_{\rm N}$ (SRO214 in-plane) and $m_{\rm S}^\perp=16m_{\rm N}$ (SRO214 out-of-plane). To incorporate the properties of its layered structure, the SRO214 Fermi surface is approximated by an ellipsoid with the cut-off angle of $\pi/10$. The magnetization $M$ and thickness $t_{\rm F}$ of the ferromagnet are set to be, respectively, $X=M/H_{\rm ex}=0.6$ and $t_{\rm F}=k_{\rm F}L=11$. We assumed the magnetization direction is parallel to the $d$-vector of the superconductor, as both magnetization of SRO113 and $d$-vector of SRO214 bulk are along the $c$-axis~\cite{Anwar2015,Maeno2012}. Figure~\ref{Calculations}(a) shows the normalized conductance as a function of the normalized bias voltage in the cases of chiral $p$-wave and non-chiral $p$-wave ($\chi = 0$) symmetry of the induced minigap in the F-layer. The barriers were taken as $Z_1=2$ and $Z_2 = 0$ (definations are given in the caption of Fig.~\ref{Calculations}).

The V-shaped induced minigap in the experimental data is reproduced well by the calculated non-chiral $p$-wave spectrum. This suggests that the induced superconductivity in the SRO113 is non-chiral $p$-wave. Calculated conductance spectra for different barrier heights, in the presence of a magnetic field and for non-zero temperatures are discussed in the Supplemental Material~\cite{SM}.
To obtain a better understanding of the proximity induced unconventional superconductivity in ferromagnets, self-consistent and/or multiband models would be needed. To gain more about the parity of the induced superconductivity, Green function model can be considered.

Now, we briefly discuss the $V_{\rm Dip}$ feature, which corresponds to the critical current of a superconducting junction \cite{Sheet2004,Yang2012}. Temperature dependence of $dI$/$dV$ (Fig. \ref{dIdV}) shows that $V_{\rm Gap}$ from Au/STO/SRO113 interface disappears around 1.1 K, as discussed above, but $V_{\rm Dip}$ persists up to the bulk $T_{\rm c}$. It indicates that $V_{\rm Dip}$ attributes to the critical current of the induced superconductivity because of the proximity effect at SRO113/SRO214 transparent interface~\cite{Sheet2004,Yang2012}. For further confirmation, we applied the Ambegaokar–Baratoff (AB) theory~\cite{AB_Theory} using the relation,

\begin{equation}
\frac{I_c(T)}{I_{c0}}=\frac{\Delta(T)}{\Delta_0}{\rm tanh}\left( \frac{\Delta(T)}{2k_BT}\right)
\end{equation}

which fits reasonably well with experimental $I_{\rm c}$($T$) data (Fig.~\ref{Analysis}(c)). 

Broken inversion symmetry at the SRO113/SRO214 interface can also induce the odd-frequency $s$-wave spin-triplet correlations, which is the only dominating pair amplitude in the diffusive system ($l_{\rm e}<t_{\rm F}$). However, in the clean limit ($l_{\rm e}>t_{\rm F}$), an odd-frequency component may coexist with the dominating even-frequency component. Such a situation can be probed with detailed differential conductance measurements at low bias, where a ZBCP may emerge within V-shaped gap. To study the proximity effect relative to the magnetization rotation of ferromagnetic layer, an F-layer with lower coercive field is desirable. Potential materials can be La$_{0.7}$Sr$_{0.3}$MnO$_3$ or La$_{0.7}$Ca$_{0.3}$MnO$_3$, if a good electronic contact (small value of $Z_2$) can be achieved with SRO214.

\section{Conclusion}

We systematically observed a superconducting induced minigap in the SRO113 ferromagnet using Au/STO/SRO113/SRO214 tunnel junctions. The minigap width roughly matches with the Thouless energy of a 15-nm thick SRO113 in a clean limit. The V-shape differential conductance around zero bias indicates the $p$-wave nature of the induced spin-triplet correlations. This is also supported by the calculations of differential conductance for non-chiral $p$-wave order parameter. Our work will pave the way towards the study of the $p$-wave spin-triplet proximity effect and play a crucial role in the development of superconducting spintronics.

\section{Acknowledgement}
We are thankful for valuable discussions with Y. Tanaka, K. Yada and A Golubov. This work is supported by the JSPS KAKENHI projects Topological Quantum Phenomena (JP22103002 and JP25103721) and Topological Materials Science (JP15H05851, JP15K21717 and JP15H05852), JSPS KAKENHI 17H04848, as well as by JSPS-EPSRC core-to-core program ''Oxide-Superspin (OSS)'' . MSA is supported as an International Research Fellow of the JSPS.

\end{document}